\documentclass[letterpaper,twocolumn]{jpsj3}
\usepackage{txfonts}
\usepackage{graphicx}
\usepackage{latexsym}
\usepackage{amsmath}
\usepackage{txfonts}
\usepackage{helvet}
\usepackage{braket}
\usepackage{amscd}
\usepackage{amssymb}
\usepackage{longtable}

\title{Site-dependent Local Spin Susceptibility and Low-energy Excitation in a Weyl Semimetal WTe$_2$}

\author{Toshiki Yokoo, Yukihiro Watanabe, Masashi Kumazaki, Masayuki Itoh, and Yasuhiro Shimizu\thanks{yasuhiro@iar.nagoya-u.ac.jp}}
\inst{Department of Physics, Nagoya University, Nagoya 464-8602, Japan} 

\abst{Site-dependent local spin susceptibility is investigated with $^{125}$Te nuclear magnetic resonance in a Weyl semimetal WTe$_2$. The nuclear spin-lattice relaxation rate $1/T_1T$ shows a dependence of the square of temperature $T$ at high temperatures, followed by a constant behavior below 50 K. The temperature dependence features Weyl fermions appearing around the linearly crossing bands. The Knight shift $K$ scales to the square root of $1/T_1T$, corroborating a predominant spin contribution in low-lying excitation. The observed dependence of $K$ and $1/T_1T$ on the four Te sites shows the site-dependent electron correlation and density of states. The angular profile of the NMR spectrum gives the anisotropic hyperfine coupling tensor, consistent with $5p$ hole occupations on Te sites. }

\begin{document}
\maketitle
\section{Introduction}
 Weyl semimetals with broken inversion symmetry display a singular electromagnetic response such as chiral anomaly and nonlinear Hall effect \cite{Xiong, Hsieh, Soluyanov, Yan, Ma}, originating from the Berry curvature in the linearly crossing energy bands. The topological phenomena suggest the presence of magnetic monopoles around the Weyl points \cite{Wu2,Feng}, which appear on the surface of the semimetals. In contrast to the surface electronic state, the low-lying excitation of the underlying bulk state is less understood owing to the lack of comprehensive magnetic measurement. Nuclear magnetic resonance (NMR) spectroscopy can be a local probe of emergent gauge fields from magnetic monopoles in Weyl semimetals \cite{Okvatovity,Dora}. It also extracts the local spin/orbital susceptibility and excitation of Weyl fermions in a site-selective manner, as extensively studied in a molecular material $\alpha$-(ET)$_2$I$_3$ \cite{Hirata} and $5d$ transition metal compounds such as TaAs and ZrTe$_5$ \cite{Yasuoka, Tian, YuLiu, Wang2, Watanabe}. The low-lying excitation measured with the nuclear spin-lattice relaxation rate $1/T_1$ sensitively depends on the energy dispersion curvature and the location of the Fermi level around the Weyl points \cite{Dora,Okvatovity, Hirosawa}.

One of the most studied materials is the quasi-two-dimensional transition-metal dichalcogenide WTe$_2$, which exhibits versatile topological phenomena such as the quantum spin-Hall effect \cite{Qian1344, Wu, Shi2, Tang}, chiral anomaly \cite{Lv}, and nonlinear Hall effect \cite{Ma}. The material also displays extremely large magnetoresistance \cite{Ali}, ferroelectricity \cite{Sharma}, and superconductivity \cite{Kang,Fatemi}. $T_d$-WTe$_2$ is classified into a type-$\rm{I\hspace{-.1em}I}$ Weyl semimetal with tilted Weyl corns owing to lattice distortion \cite{Soluyanov}. The charge compensation of electron and hole Fermi pockets under magnetic field \cite{Wang, Pletikosi} causes the nonsaturating magnetoresistance in a clean sample \cite{Ali,Pletikosi}, where spin-orbit coupling \cite{Jiang} and Lifshitz transition \cite{Wu3,Zhu} also play a crucial role in the charge compensation. 
	\begin{figure}
	\includegraphics[scale=0.5]{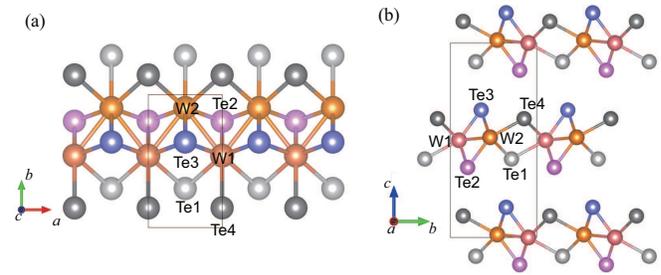}
	\caption{\label{Fig:crystal} 
(Color online) Crystal structure of $T_d$-WTe$_2$ composed of two tungsten (W1, W2) and four tellurium (Te1, Te2, Te3, Te4) sites in the nonsymmorphic $Pmn2_1$ lattice. The W$_2$Te$_4$ layer in the $ab$ plane (a) is stacked along the $c$ axis. Two equivalent atomic sites in the unit cell are related by the glide symmetry in the $bc$ plane (b) but become inequivalent as the magnetic field is tilted from the mirror plane along the $ab$ and $ca$ planes.   
	}
	\end{figure}

The crystal structure of WTe$_2$ (orthorhombic space group $Pmn2_1$) is composed of the W$_2$Te$_4$ layers stacked along the $c$ axis, as shown in Fig. \ref{Fig:crystal} \cite{Rano}. There are crystallographically inequivalent two tungsten (W1, W2) and four tellurium sites (Te1, Te2, Te3, Te4), and two each are included in the unit cell. Each Te site has the same site symmetry and three coordinations to W atoms. Owing to the large distortion along the $a$ axis, there is direct W--W bonding along the $a$ axis, and the W--Te bond length of the Te2/Te3 sites ($\sim 2.7$\AA) is shorter than that of the Te1/Te4 sites ($\sim 2.8$\AA). The difference is expected to give a different partial density of states and orbital occupation, which are observed as the local spin susceptibility and the anisotropic hyperfine coupling, respectively. In the band structure, the Weyl points are located near the $\Gamma$ point of the Brillouin zone \cite{Augustin}. Both W $5d$ and Te $5p$ bands significantly contributes to the Fermi level. Therefore, $^{125}$Te NMR spectroscopy can be used as a local probe of the electronic state of the Weyl semimetal.

	In this paper, we conduct site-selective $^{125}$Te NMR measurements in WTe$_2$ single crystals. The local susceptibility observed by the Knight shift gives the site-dependent density of states contributing to the Weyl points. The result compensates the $k$-space surface information probed by optical \cite{Kimura, Frenzel, Guan} and photoemission spectroscopy \cite{Bruno, Feng, Wu2, Tang, Wu3}. On the basis of crystal structure data, the NMR spectra are assigned to four Te sites with the distinct local spin susceptibility. We evaluate the site-dependent Te $5p$ orbital electron/hole occupation from the Knight shift tensor. 
	
\section{Experimental Method}
	Single crystals of WTe$_2$ were prepared by two methods: chemical vapor transport and Te-flux methods. The latter is expected to give higher quality stoichiometric crystals \cite{Ali_2015,Zhu}. In the chemical transport method, we start from powdered tungsten and tellurium kept at 700$^\circ$C in a vacuum quartz tube for two days and then at 750$^\circ$C for another two days after grinding \cite{Ali}. Crystals were grown in a three-zone furnace with a transport agent Br$_2$ and a temperature gradient of 750--650$^\circ$C for one week. In the self-flux method, WTe$_2$ and tellurium powder were also heated to 700$^\circ$C for 2 days and 750$^\circ$C for another 2 days. After maintaining at 900$^\circ$C for 10 h, the temperature decreased to 460$^\circ$C for 100 h. The Te flux was separated by centrifugation and then evaporated in a three-zone furnace with a temperature gradient of 465-280-250$^\circ$C for two days. 
	
    The crystals (named \#S1 and \#S2) obtained by the transport and flux methods have dimensions of 5 mm $\times$ 0.3 mm $\times$ 0.05 mm and 5 mm $\times$ 1 mm $\times$ 0.2 mm, respectively. For NMR measurements of \#S1, several thin plate crystals were aligned and stacked. Magnetic susceptibility was measured with a magnetometer (MPMS-XS, Quantum Design Ltd.) at 7 T along the crystal axes. 
    We obtained $^{125}$Te NMR spectra by Fourier transformation of spin-echo signals after the rf pulses $t_{\pi/2}-\tau-t_{\pi}-\tau$ (duration $t_{\pi/2} = 1$ $\mu$s and the interval time $\tau = 20 - 30$ $\mu$s) under constant magnetic fields $H = 7.389$ and 9.0891 T. $^{125}$Te Knight shift $K$ was defined by the relative frequency shift $K = (\nu - \nu_0)/\nu_0$ using the bare resonance frequency $\nu_0 = \gamma_nH$ for the nuclear gyromagnetic ratio $\gamma_n = 13.454$ T/MHz \cite{Robin}.
	
\section{Experimental Results}

	\begin{figure*}[h]
	\includegraphics[scale=0.9]{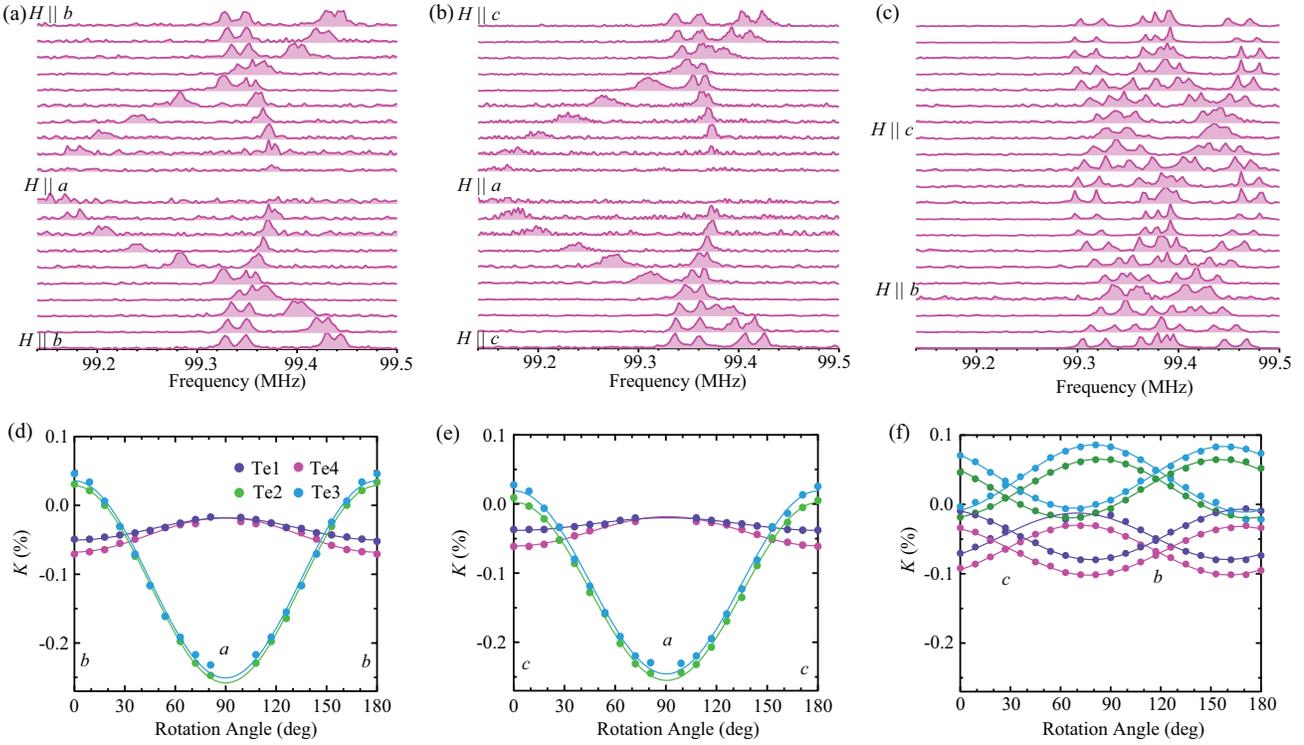}
	\caption{\label{Fig:shift} 
(Color online) $^{125}$Te NMR spectrum and Knight shift $K$ for a single crystal of WTe$_2$ (\#S2) at 200 K. The spectrum was measured with $9^\circ$-step rotation of the crystal along (a,d) $c$, (b,e) $b$, and (c,f) $a$ axes using a two-axis rotator. The absence of data around the $a$ axis is due to the alignment of the magnetic field parallel to the rf field. The curves are fitting results using Eq. (1).
	}
	\end{figure*}

    The angular dependence of $^{125}$Te NMR spectra at 200 K for a single crystal of the sample \#S2 is shown in Fig. \ref{Fig:shift} for a sample rotated along the orthogonal crystal axes. Four resonance lines observed at $b$ and $c$ axes consist of two doublets exhibiting distinct anisotropy for the $ab$- and $ca$-plane rotations. One doublet shows a large negative shift with a minimum at the $a$ axis. Another shows a weak dependence on the field direction with a maximum at the $a$ axis. As the magnetic field is rotated in the $bc$ planes, each spectrum splits into two because two Te sites related by glide symmetry in the unit cell become inequivalent against the magnetic field [See Fig. \ref{Fig:shift}(c)].
    
    We refer to the crystal structure for the spectrum assignment. Namely, four Te sites are classified into two groups in terms of the W--Te bond length (Te1/Te4 and Te2/Te3) and possess the identical site symmetry (1a) and the same coordination number (three) to W ions. The stronger hybridization to W bands is expected for Te2/Te3 with shorter W--Te bonds than those for Te1/Te4, resulting in a larger spin density proportional to the amplitude of the Knight shift anisotropy. Thus, we assigned the NMR spectrum into four Te sites, as shown in the symbols of Fig. 2(d). 

   The angular dependence of the $^{125}$Te Knight shift $K$ obtained from a relative frequency shift of each NMR spectrum is shown in Figs. 2(d)--2(f). The anisotropy is governed by the dipolar hyperfine interaction with the Te $5p$ spins, while the isotropic part comes from the Fermi contact interaction and the diamagnetic constant terms. The site dependence of $K$ reflects the local spin susceptibility and the $5p$ orbital occupation at the Te sites with the different densities of states. The angular dependence of $K$ is analyzed with a sinusoidal function \cite{Slichter}  
	\begin{eqnarray}
	K = k_1 {\rm cos}2\theta + k_2 {\rm sin}2\theta + k_3 
	\label{Eq5}
	\end{eqnarray} 
for each rotation along the crystal axis with the fitting parameters ($k_1$, $k_2$, $k_3$), as shown in Figs. 2(d)--2(f). For the $bc$-plane rotation, for example, the fitting gives the components of the Knight shift tensor {\sf K} using the relations, $k_1 = (K_{bb} + K_{cc})/2$, $k_2 = (K_{bb} - K_{cc})/2$, and $k_3 = K_{bc} = K_{cb}$. 

From the mirror reflection symmetry along the $ab$ and $ca$ planes, the $a$ axis must be one of the principal axes of ${\sf K}$, where the off-diagonal components $K_{ab}$ and $K_{ac}$ vanish. Thus, we can express ${\sf K}$ as 
    	\begin{equation}
	    {\sf K} =
	\begin{pmatrix}
    K_{aa} & 0 & 0 \\
    0 & K_{bb} & K_{bc} \\
    0 & K_{cb} & K_{cc} \\
    \end{pmatrix}.
	\end{equation}
After the diagonalization of the tensor, we obtained the principal components of the Knight tensor $(K_{XX}, K_{YY}, K_{ZZ})$ for four Te sites, as listed in Table \ref{tab:Te1Te4}. The anisotropy reflects the Te $5p$ orbital occupation dominating the local spin/orbital susceptibility, as discussed below. 

\begin{table}
	\centering
	    \begin{tabular}{cccccccc} 
\hline
& $K_{aa}$ & $K_{bc}$ & $K_{bb}$ & $K_{cc}$ & $K_{XX}$ & $K_{YY}$ & $K_{ZZ}$\\ \hline \hline 
Te1   & -0.018 & 0.036 & -0.047 & -0.042  & -0.008 & -0.018 & -0.081 \\ 
Te4   & -0.018 & 0.036 & -0.064 & -0.068  & -0.030 & -0.018 & -0.103 \\
Te2   & -0.258 & 0.039 & 0.042 & 0.004  & 0.042 & 0.004 & -0.258 \\ 
Te3   & -0.251 & 0.045 & 0.055 & 0.022  & 0.086 & -0.009 & -0.251 \\ \hline
	    \end{tabular}
	    \caption{Components of $^{125}$Te Knight shift tensors {\sf K} (\%) for four Te sites in Eq. (1), obtained from the angular dependence measurement at 200 K in Fig. 2. The principal axes $(X, Y, Z)$ for the diagonalized Knight shift tensor correspond to the directions where $K$ shows a maximum ($X$) and a minimum ($Z$). }
	    \label{tab:Te1Te4}
	    \end{table}

	\begin{figure*}[h]
	\begin{center}
	\includegraphics[scale=0.8]{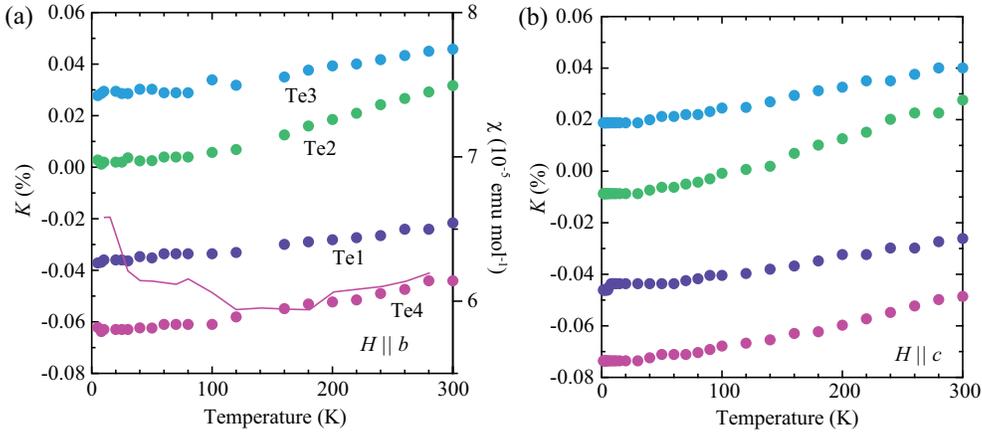}
	\caption{\label{Fig:shifttemp} 
(Color online) Temperature dependence of $^{125}$Te Knight shift $K$ for four Te sites under the magnetic field along the $b$ axis (a) and the $c$ axis (b) in the single crystal of WTe$_2$ (\#S2) at 9.083 T. The bulk magnetic susceptibility (a solid curve) in the right-hand axis in (a) was measured at 7.0 T, where the core diamagnetic susceptibility was subtracted. 
	}
	\end{center}
	\end{figure*}
	
	Figure \ref{Fig:shifttemp} shows the temperature dependence of $K$ along the $b$ and $c$ axes. Upon cooling, $K$ linearly decreases to 50 -- 100 K and becomes nearly constant at low temperatures, scaling to the bulk magnetic susceptibility $\chi$ above 100 K. The behavior is characteristic of Dirac/Weyl semimetals having a linear band dispersion \cite{Yasuoka, Wang2, Tian}. The crossover temperature from the linear to constant behavior reflects the energy scale of chemical potential crossing the Weyl corn. At low temperatures, the system becomes a weakly correlated metal with a finite density of states. 
	
The temperature dependence of $K$ or the spin susceptibility $\chi_{\rm s}$ has been calculated theoretically for three-dimensional (3D) Weyl semimetals \cite{Okvatovity}. For the density of states linear to energy, $\rho(E) \propto E$, in Dirac/Weyl semimetals, $\chi_{\rm s}$ decreases linearly with $k_{\rm B}T$, which is distinct from the $T$-invariant behavior in normal metals. $K$ also includes the component of the orbital susceptibility $\chi_{\rm orb}$, which is sensitive to the chemical potential. In comparison with the calculation, the downward Knight shift suggests a negative chemical potential for the Te-dominant band \cite{Okvatovity}, consistent with the hole-like carrier expected from the band calculation in WTe$_2$ \cite{Augustin}.

As for the site dependence, one of the Te sites in the doublet pair (Te2 and Te4) displays a stronger thermal variation of $K$. As a result, the doublet splitting increases upon cooling and reaches $\sim 0.03\%$ for $H || b$ and $c$. Here the Te2/Te4 sites are bonded to two W1 among three W--Te bondings, while Te1/Te3 are connected to two W2. The coordination difference can lead to the site-dependent curvature of the Te bands, leading to a thermal change in local spin susceptibility. The result should be confirmed by the band structure calculation taking into account the atomic sites. We obtained the net hyperfine coupling constants $H_{\rm hf}$ (= 22--26 T/$\mu_{\rm B}$) for Te sites as the linearity coefficient of the $K$--$\chi$ plot, which are comparable to that of ZrTe$_5$ ($\sim 15$ T/$\mu_{\rm B}$) \cite{Tian2} and highlight the predominant onsite hyperfine interaction.  
	
	\begin{figure*}[h]
	\begin{center}
	\includegraphics[scale=0.9]{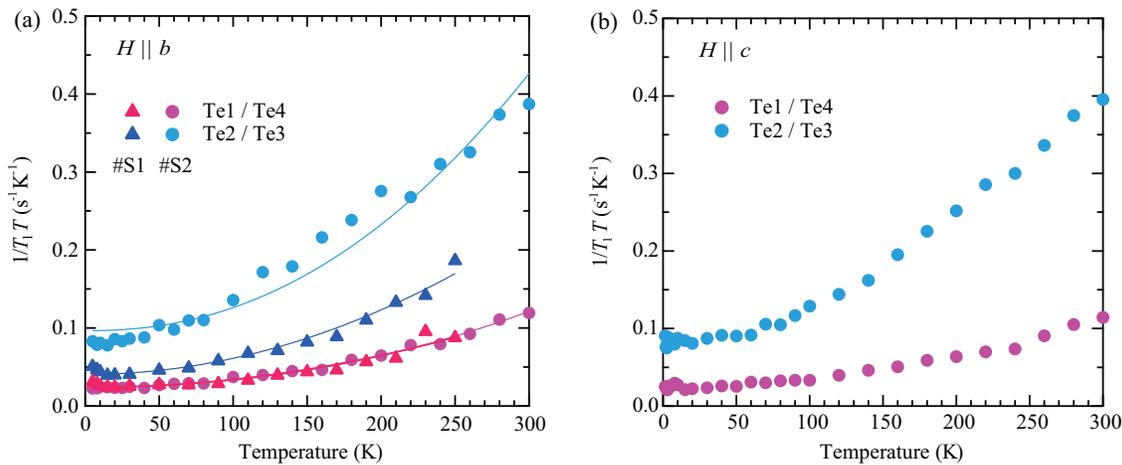}
	\caption{\label{Fig:T1} 
(Color online) (a) Temperature dependence of $^{125}$Te nuclear spin-lattice relaxation rate divided by temperature, $1/T_1T$, for the single crystals (\#S1 and \#S2) of WTe$_2$ under magnetic field along the $b$ axis. The solid curve is the model fitting for Weyl fermions. (b) $1/T_1T$ for \#S2 measured along the $c$ axis. 
	}
	\end{center}
	\end{figure*}
    
The temperature dependence of the nuclear spin-lattice relaxation rate divided by $T$, $1/T_1T$, is measured under magnetic field along the $b$ and $c$ axes, as shown in Fig. \ref{Fig:T1}. Here, the $T_1$ values for two Te sites within the doublet (Te1 or Te4, Te2 or Te3) are not distinguishable because they are almost the same. Interestingly, the sample dependence was observed for the Te2/Te3 site, which is considered to be sensitive to chemical potential. For sample \#S2, $1/T_1T$ for Te2/Te3 is about three times larger than that for Te1/Te4, indicating a higher spin density for Te2/Te3, consistent with the result of $K$. We find that $1/T_1T$ is strongly suppressed with decreasing temperature, approximately following a parabolic function $\sim C_1T^2 + C_2$. The constant term indicates the residual density of state, characteristic of the type-II Weyl semimetal. This is consistent with the observation of the small Fermi surface by photoemission spectroscopy \cite{Wu3, Sante}. Similar $1/T_1T$ behavior has been commonly observed in type-I Weyl \cite{Yasuoka, Wang2} and 3D Dirac semimetals \cite{Suetsugu, Watanabe} with the residual density of states. 

The theoretical formula of $1/T_1T$ for 3D Dirac fermions is expressed as
\begin{equation}
\begin{array}{cc}\label{eq:4}
\begin{split}
    \displaystyle\frac{1}{T_1T} &= \displaystyle\frac{2\pi}{3}\mu_0 ^2\gamma_n ^2 e^2c^{*4}\\ & \times \displaystyle\int_{-\infty}^{\infty}dE\left[-\displaystyle\frac{\partial f(E,\mu)}{\partial E}\right]\displaystyle\frac{\rho^2(E)}{E^2}{\rm ln}\displaystyle\frac{2(E^2-\Delta^2)}{\omega_0|E|} , 
\end{split}
\end{array}
 \end{equation}
where $\rho(E)$ is the density of states, $f(E,$ $\mu)$ the Fermi distribution function, $c^* = (\Delta /m^*)^{1/2}$ the velocity with the effective mass $m^*$, $\omega_0$ the NMR frequency, and $\Delta$ half the band gap \cite{Hirosawa, Maebashi}. $1/T_1T$ scales to $\sim T^2$ for $\mu$, $\Delta \ll T$ and then approaches a constant value for a temperature range $T \ll \mu$, following Korringa's law in normal metals. Although the theoretical calculation does not assume the strong tilting of the Weyl corns, our experimental result well fits to the theoretical calculation of $1/T_1T$, as shown in Fig. 4(a). The agreement suggests that the low-lying excitation is governed by the band curvature around the Weyl points and the density of states, which is commonly linear in energy for both the type-I and type-II Weyl semimetals. Under high magnetic fields, the continuous band can be squeezed into the Landau levels. However, we did not observe indication of the quantization and suppression of the spin density by charge compensation in the field range of 5--9 T. 
    
\begin{figure}[h]
\begin{center}
    \includegraphics[scale=0.5]{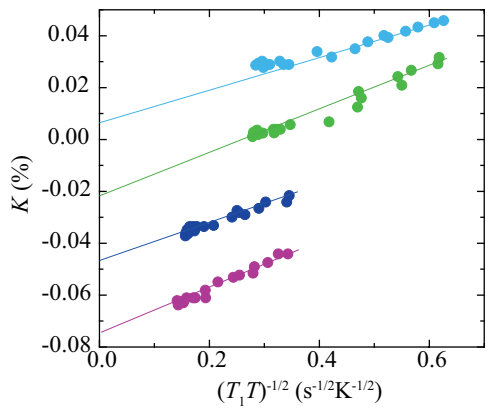}
	\caption{\label{Fig:T1TK} 
(Color online) $^{125}$Te Knight shift $K$ plotted against $(1/T_1T)^{1/2}$ as an implicit function of temperature in WTe$_2$ (\#S2).
	}
	\end{center}
	\end{figure}

In weakly correlated metals, $(1/T_1T)$ is proportional to the square of the density of states, $\rho(E)^2$, at the Fermi level, while $K$ directly scales to $\rho(E)$ depending on temperature. Following the extended Korringa's relation, $K$ linearly scales to $(1/T_1T)^{1/2}$ \cite{Wang2}. Indeed, we find a good linear relationship between $K$ and $(1/T_1T)^{1/2}$ for an extensive temperature range from 300 to 10 K, as shown in Fig. \ref{Fig:T1TK}. Here, the $y$-offset extracts the $T$-independent term governed by the diamagnetic chemical shift due to inner electrons and the paramagnetic Van-Vleck shift. 

The linearity yields the ratio between $1/T_1T$ and $K^2$, which is called the Korringa ratio defined by $\mathcal{K}(\alpha )$ = $(T_1TK^2)^{-1}/S_0$, where $S_0 = \hbar(\gamma_e/\gamma_n)^2/(4\pi k_{\rm B}) = 2.64 \times 10^{-6}$ sK is the Korringa constant for free electrons with the gyromagnetic ratio $\gamma_e$. From the linearity in Fig. \ref{Fig:T1TK}, we obtained $\mathcal{K}(\alpha )$ as 2.6--5.5 depending on the Te site. The enhancement factor can be a measure of the site-dependent electron correlation of Weyl fermions \cite{Isobe} and orbital fluctuations \cite{Okvatovity,Dora}. In the present experiment, we neglect the anisotropy of the hyperfine coupling because of the lack of the $a$-component of $K$, which may cause the overestimation of $\mathcal{K}(\alpha )$. 

\section{Discussion}
We discuss the Te $5p$ orbital occupation in terms of the anisotropic hyperfine coupling obtained from the Knight shift measurement. The band structure of WTe$_2$ has been intensively studied on the basis of the density functional theory and the local density approximation \cite{Augustin, Sante, Ali}. There are two Weyl points and Fermi pockets consisting of the electron and hole-like bands. Not only tungsten $5d$ bands but also tellurium $5p$ bands have a significant weight on the Weyl points carrying the low-lying excitation \cite{Augustin}. However, the energy bands are not labeled with crystallographically inequivalent atomic sites. 

The anisotropic part of the Knight shift comes from the magnetic dipolar hyperfine interaction with $5p$ spins. The different anisotropy of the hyperfine coupling tensor suggests the site-dependent $5p$ orbital occupation ($p_x$, $p_y$, $p_z$) at the Fermi level. The dipolar hyperfine coupling tensors of $p_x$, $p_y$, and $p_z$ orbitals have the diagonal components proportional to ($2$, $-1$, $-1$),  ($-1$, 2, $-1$), and ($-1$, $-1$, 2), respectively \cite{Abragam}. 

The observed $K$ in Fig. 2 shows a prominent minimum for Te2/Te3 around the $a (x)$ axis. The large negative $K_{\rm ZZ}$ indicates the doubly degenerate $p_y$ and $p_z$ electron or the $p_x$ hole occupation near the Fermi level. In contrast, $K$ for Te1/Te4 has a minimum in the $bc$ plane parallel to the Te1--Te4 bond direction, suggesting a predominant $p_y$ or $p_z$ hole occupation. The amplitude of the Knight shift is suppressed because of the lower density of states. The orbital selective property of the band structure has recently been investigated via photoexcitation \cite{Guan}, and the contribution of Te $5p$ orbitals to the hole-like bands crossing the Fermi level has been shown, consistent with the result of Knight shifts. Our experimental result of site-selective NMR spectroscopy suggests that all the Te sites contribute to the low-energy property in the orbital-dependent manner. 

\section{Conclusion}
Local spin and orbital susceptibility was investigated via $^{125}$Te NMR measurements on single crystals of the Weyl semimetal WTe$_2$. We obtained the local spin susceptibility of the four Te sites that are classified into two groups with different anisotropies depending on the W--Te bond length. The temperature dependence of the nuclear spin-lattice relaxation agrees with the theoretical calculation in Weyl semimetals. We confirmed Korringa's law in an extensive temperature range, consistent with the weakly correlated metal. The enhancement factor of the Korringa constant suggests orbital fluctuations of Weyl fermions. The predominant hole occupation in $p$ orbitals was evaluated from the anisotropy of the Knight shift. These results complement the results of optical and photoemission spectroscopy. 

$Note$: Similar $1/T_1$ data of WTe$_2$ for powder samples were found in the literature during the preparation of the manuscript, which qualitatively agree with our data \cite{Antonenko, papawassiliou}. 

\section*{Acknowledgements}
We thank S. Inoue and T. Jinno for technical support. This work was supported by JSPS KAKENHI (Grants No. JP19H01837, JP16H04012, and JP19H05824).


\end{document}